\documentclass[cits]{PoS}
\usepackage{xspace}
\newcommand{\Al}{$^{26}$Al\xspace}

\newcommand{\degree}{$^{\circ}$}
\newcommand{\Fe}{$^{60}$Fe\xspace}

\let\jnl=\rmfamily
\def\refe@jnl#1{{\jnl#1}}%
\newcommand\aj{\refe@jnl{AJ}}%
\newcommand\actaa{\refe@jnl{Acta Astron.}}%
\newcommand\araa{\refe@jnl{ARA\&A}}%
\newcommand\apj{\refe@jnl{ApJ}}%
\newcommand\apjl{\refe@jnl{ApJ}}%
\newcommand\apjs{\refe@jnl{ApJS}}%
\newcommand\ao{\refe@jnl{Appl.~Opt.}}%
\newcommand\apss{\refe@jnl{Ap\&SS}}%
\newcommand\aap{\refe@jnl{A\&A}}%
\newcommand\aapr{\refe@jnl{A\&A~Rev.}}%
\newcommand\aaps{\refe@jnl{A\&AS}}%
\newcommand\azh{\refe@jnl{AZh}}%
\newcommand\memras{\refe@jnl{MmRAS}}%
\newcommand\mnras{\refe@jnl{MNRAS}}%
\newcommand\na{\refe@jnl{New A}}%
\newcommand\nar{\refe@jnl{New A Rev.}}%
\newcommand\pra{\refe@jnl{Phys.~Rev.~A}}%
\newcommand\prb{\refe@jnl{Phys.~Rev.~B}}%
\newcommand\prc{\refe@jnl{Phys.~Rev.~C}}%
\newcommand\prd{\refe@jnl{Phys.~Rev.~D}}%
\newcommand\pre{\refe@jnl{Phys.~Rev.~E}}%
\newcommand\prl{\refe@jnl{Phys.~Rev.~Lett.}}%
\newcommand\pasa{\refe@jnl{PASA}}%
\newcommand\pasp{\refe@jnl{PASP}}%
\newcommand\pasj{\refe@jnl{PASJ}}%
\newcommand\skytel{\refe@jnl{S\&T}}%
\newcommand\solphys{\refe@jnl{Sol.~Phys.}}%
\newcommand\sovast{\refe@jnl{Soviet~Ast.}}%
\newcommand\ssr{\refe@jnl{Space~Sci.~Rev.}}%
\newcommand\nat{\refe@jnl{Nature}}%
\newcommand\iaucirc{\refe@jnl{IAU~Circ.}}%
\newcommand\aplett{\refe@jnl{Astrophys.~Lett.}}%
\newcommand\apspr{\refe@jnl{Astrophys.~Space~Phys.~Res.}}%
\newcommand\nphysa{\refe@jnl{Nucl.~Phys.~A}}%
\newcommand\physrep{\refe@jnl{Phys.~Rep.}}%
\newcommand\procspie{\refe@jnl{Proc.~SPIE}}%

\title{Massive-Star Nucleosynthesis and INTEGRAL}

\ShortTitle{GRIPS}

\author{\speaker{Roland Diehl}\\
        Max Planck Institut f\"ur extraterrestrische Physik, D-85748 Garching, Germany\\
        E-mail: \email{rod@mpe.mpg.de}}

\abstract{Products from massive-star nucleosynthesis have been measured with SPI on INTEGRAL: Characteristic gamma-ray lines from radioactive decays of long-lived $^{26}$Al and $^{60}$Fe isotopes, and from $^{44}$Ti decay ($\tau$=89y). Detections of both these isotopes has laid the foundation to peek into massive-star interiors, through different views at those measurements. The gamma-ray flux can be converted into amounts of these radioactive isotopes, but the constraints which derive from this for massive star models involve additional steps. -- 
Earlier results had demonstrated the basic constraints inherent to such radioactivity data, i.e. detection of $^{26}$Al is a calibration for massive-star yields, $^{60}$Fe enables an isotopic-yield ratio test eliminating modeling and observing bias aspects, and $^{44}$Ti searches showed that its production does not occur homogeneously over core-collapse events. The current status of $^{26}$Al and $^{60}$Fe observations and their analysis is reaching a threshold for astrophysical insights: Specific regions of massive star groups and their radioactivity gamma-rays have recently been investigated, such as Sco-Cen and Orion. As for $^{44}$Ti, spectroscopical information constrains ejection velocities of these inner supernova ejecta for the Cas A event. These parameters for massive stars can be related to  other fields of astronomy, of nuclear physics experiments and theory, and of theoretical astrophysics. Thus,  model improvements appear on the horizon, as e.g.  implemented in the EuroGenesis program framework of the European Science Foundation.}

\FullConference{8$^{th}$ INTEGRAL Science Workshop\\
		{\it The Restless Gamma-Ray Universe} \\
		 October 27-30, 2010\\
		 Dublin, Ireland}

\begin{document}

\section{INTEGRAL and Massive-Star Issues} 
INTEGRAL made several contributions which relate to massive stars and their science: Deeply-embedded sources of gamma-ray emission in the 100~keV domain were discovered, flares were discovered from massive-star binaries, and  diffuse radioactivity from $^{26}$Al and $^{60}$Fe were measured. Different aspects of massive stars are addressed, from their parental molecular clouds and the spatial distribution of massive stars in the Galaxy, through the properties of their wind, to the nuclear burning processes in their interiors. Here we discuss only the latter aspect, the $^{26}$Al and $^{60}$Fe measurements and what they tell us about massive-star interiors; additionally, we discuss how radioactive $^{44}$Ti data from Cas A may help to understand the final core collapse of massive stars.

Like all stars, nuclear burning and the associated energy release is the origin of stellar stability against the gravitational contraction. But in evolution of massive stars after their main sequence, i.e. core-hydrogen burning phase, nuclear burning occurs in the stellar core and also in different shells, with energy and matter transport between. The nuclear processes in such burning shells hence depend on the previous stellar evolution and isotopic changes resulting from this, and those transport processes. Moreover, stability is not always a realistic assumption, as stellar evolution accelerates in late stages, and the hydrodynamic adjustments of matter throughout the star cannot be assumed to have occurred, but may proceed during significant parts of the shell-burning stages of evolution. Therefore, evolution is complex and uncertain: Magnetic and stellar rotation effects add complexity, and may be important. Finally, the neutrino process in the explosive nuclear burning during the core collapse event may significantly affect massive-star yields for the isotopes discussed.  Therefore, it is valuable to assemble the different types of astronomical information which can be obtained to enlighten astrophysical processes in massive-star interiors, beyond what observations of their surfaces may be able to tell us. 

\begin{figure}
\centering
\includegraphics[width=0.48\textwidth]{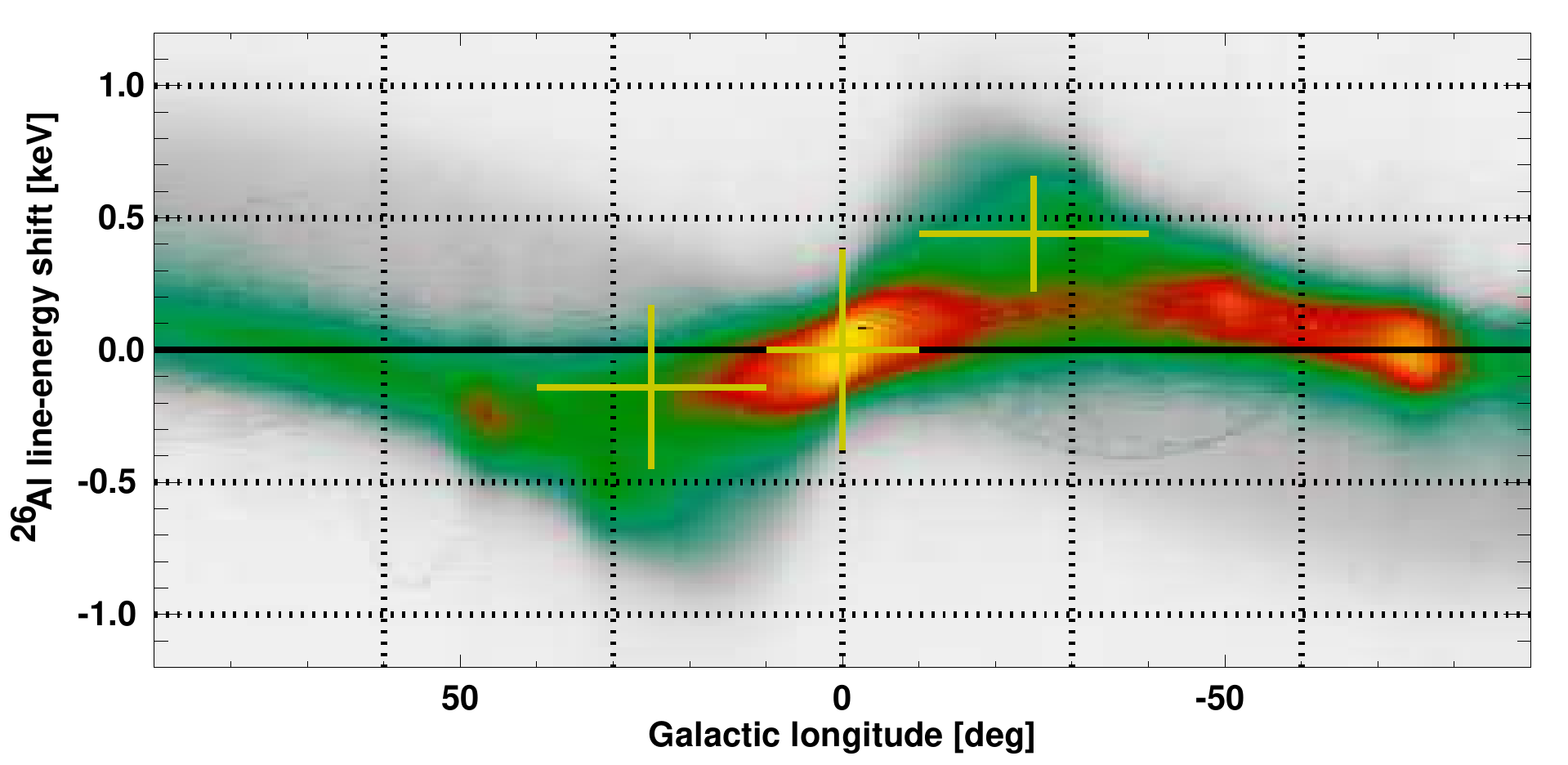}
\includegraphics[width=0.48\textwidth]{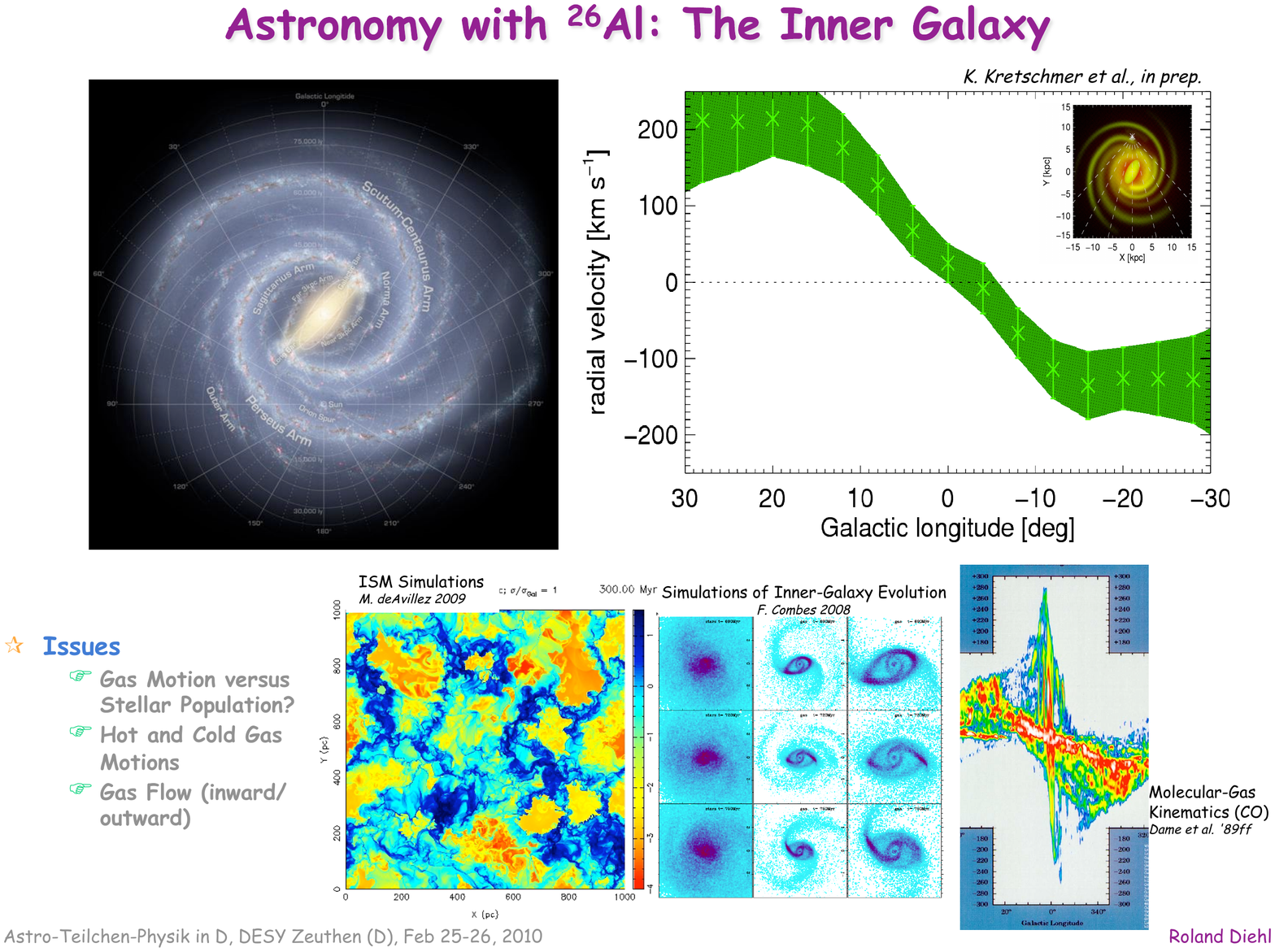}
\caption{The early hints for systematic Doppler-shifts of the $^{26}$Al gamma-ray line as expected from large-scale galactic rotation (left) were now greatly improved from refined analysis and deeper exposures (right). This establishes that sources throughout the inner parts of the Galaxy contribute; yet, it also hints at deviations of the rotational behavior of hot interstellar gas around massive stars from homogenous differential (in radius) and symmetric (in azimuth) galactic rotation pattern. (see text for explanation of details).}
\label{fig:al_rotation-signature}
\end{figure}

Earlier observations of the sky in gamma-rays from radioactive by-products of massive-star nucleosynthesis had provided foundations of their astrophysics discussions: (1) The detection of $^{26}$Al  and its imaging had established the possibility to interpret these data as a calibration for massive-star yields, using an assumed spatial distribution model and a mass distribution function of massive stars. From the integrated brightness of the Galaxy in $^{26}$Al, thus it seemed that both the supernova stage and the wind phases in the Wolf-Rayet phases contribute significantly to the massive star yield in this isotope. With better statistics and the (modest, about 3\degree) imaging resolution, the SPI/INTEGRAL data  showed that the line centroid appeared to shift by a few tenths of keV with Galactic longitude as expected from the large-scale rotation of the Galaxy, which was taken to support the assumption of measuring the Galaxy-wide contribution of $^{26}$Al sources  (see Fig.~\ref{fig:al_rotation-signature}, \cite{2006Natur.439...45D}). (2) The detection of $^{60}$Fe gamma-rays attributed to the same sources, i.e. massive stars through their late-phase shell-burning nucleosynthesis, enables an isotopic-yield ratio test eliminating modeling and observing bias aspects: In determination of an integrated isotopic-yield ratio for $^{60}$Fe/$^{26}$Al biases in source distances or observational detection efficiencies would cancel out. Thus, the integrated yield ratio can be compared quantitatively and more precisely with the observed Galaxy-integrated brightness ratio. (3) $^{44}$Ti searches showed that its production does not occur homogeneously over core-collapse events, and that the Cas A supernova event may have produced an exceptionally-high amount of $^{44}$Ti, possibly related to asphericity of the supernova explosion. 

We must be aware, however, that those earlier results do include (as always) several assumptions about massive-star properties in our Galaxy, which are taken from our astrophysical knowledge base, yet may be misguided in important details. Specifically, using Galaxy-integrated massive-star yields for astrophysical interpretation does not cater for the likely different evolutionary states of massive-star groups across the Galaxy, which might lead to significant deviations from the steady-state assumption inherent to such large-scale averaging. The $^{26}$Al (and possibly also $^{60}$Fe) observations and their analysis are beginning to reach a threshold for evaluating specific regions of massive star groups, wherein the local situation and history of star formation can be included, to quantitatively tighten comparisons between expectations and observations because adjusted to richness, age, and distance of specific groups of stars.

\section{Recent Results} 

The extension of the INTEGRAL mission and its deepening of exposure along the extended plane of our Galaxy has allowed the independent detections of $^{26}$Al emission and measurements of their brightness, line width, and line centroids: 
The systematic Doppler-shifts of the $^{26}$Al gamma-ray line as expected from large-scale galactic rotation are shown in Fig.~\ref{fig:al_rotation-signature}. In the left graph the color scale represents expectations from a standard rotation pattern and source locations following the Galactic CO distribution, with data points for three different lines of sight). These constraints were now greatly improved from refined analysis (more data points along Galactic longitude) due to deeper exposures (right). This establishes that sources throughout the inner parts of the Galaxy contribute; yet, it also hints at deviations of the rotational behavior of hot interstellar gas around massive stars from homogenous differential (in radius) and symmetric (in azimuth) galactic rotation pattern. Note that the ordinates are scaled differently in these graphs, between the observationally-observed line shift (left) and the inferred bulk gas velocity along a line of sight (right).

\begin{figure}  
\centering 
\includegraphics[width=0.7\textwidth]{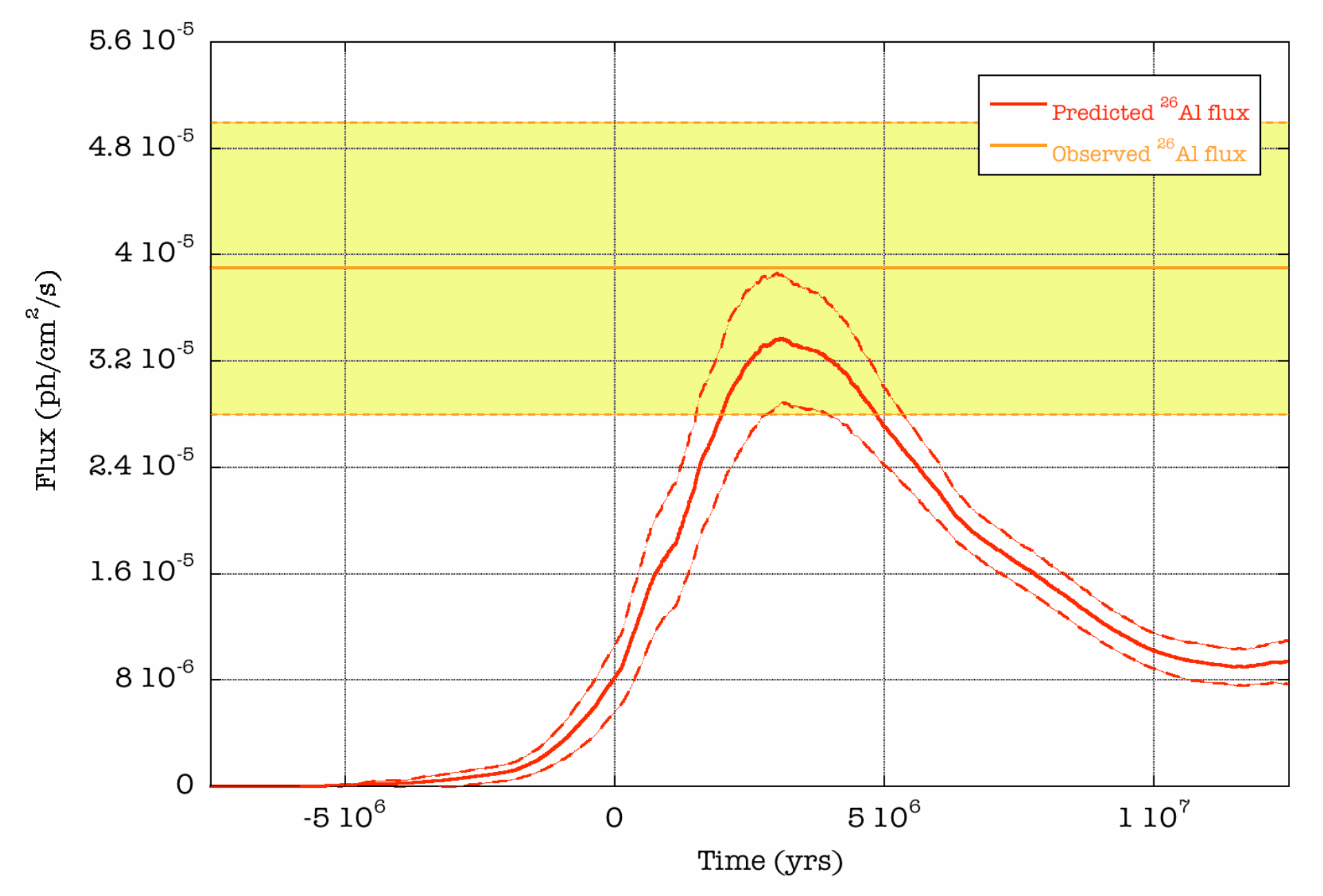}
\caption{The time history of \Al production in the Cygnus complex, as compared to the gamma-ray observations. Expectations from such populations synthesis are on the low side of observed \Al gamma-rays, though consistent within uncertainties. The shaded area presents the range given by the \Al gamma-ray data, the dashed lines bracket the uncertainty range of predictions from recent massive-star models through population synthesis. \cite{2010A&A...511A..86M}.
}
\label{fig:al_cygnus} 
\end{figure}   

For the Cygnus region, a re-analysis of the stellar populations and the groups of stars that are candidate sources of observed $^{26}$Al, $^{60}$Fe, and positrons (from $^{26}$Al decay) has been made, and converted into estimated brightness for the respective gamma-ray lines (Martin et al. 2010). Along this line of sight, there are 6 prominent OB associations at distances ranging from 0.7 to 2.5~kpc \cite{2002NewAR..46..535P}, but it appears that the Cygnus OB2 association dominates by far the stellar census of this complex. The age and distance of Cyg OB2 is 2.5~My and 1.57~kpc, respectively.  Because of this young age, contributions from core-collapse supernovae to \Al production should be small or absent, while Wolf-Rayet-wind ejected \Al from hydrostatic nucleosynthesis may be assumed to dominate. Thus \Al gamma-rays from the Cygnus region could disentangle the different \Al production phases and regions within the same massive stars: In galactic-averaged analysis, one assumes a \emph{steady state} situation of \Al decay and production, such that the complete age range of stars is represented and contributes to \Al production with its time-averaged numbers of stars per age interval and their characteristic \Al ejection from either process (hydrostatic, or late-shell burning plus explosive; \cite{2006ApJ...647..483L}).  Here, the age of currently-ejecting massive star groups suggests that \Al production is predominantly due to Wolf-Rayet wind ejection.
Comparison to observations indicates that in general the $^{26}$Al observations are met and both $^{60}$Fe and positrons should remain below SPI's sensitivity (Fig.~\ref{fig:al_cygnus}). If the locally-adjusted (lower) metallicity of the Cygnus region, however, is applied in the respective population synthesis brightness prediction, it appears that more $^{26}$Al is seen that expected, and that future data with deeper exposures may have a chance to detect $^{60}$Fe, being brighter at lower metallicity, in comparison. 
Additionally, for a young and active region of massive-star action, one may plausibly assume that the interstellar medium would be peculiar and probably more dynamic than in a large-scale average.  INTEGRAL's early  hints for a broadened \Al gamma-ray line now turn out to not hold up, and the \Al line seen from the Cygnus region is compatible with the laboratory energy (i.e. no bulk motion exceeding tens of km~s$^{-1}$) and with instrumental line width (i.e. no excessive Doppler broadening beyond $\sim$200~km~s$^{-1}$ \cite{2009A&A...506..703M}. Note that \Al ejection from Wolf Rayet winds would be with $\sim$1500~km~s$^{-1}$, decelerating as circumstellar gas would be swept up.
These results are discussed in detail in \cite{2010A&A...511A..86M}.

\begin{figure}  
\centering 
\includegraphics[width=0.6\textwidth]{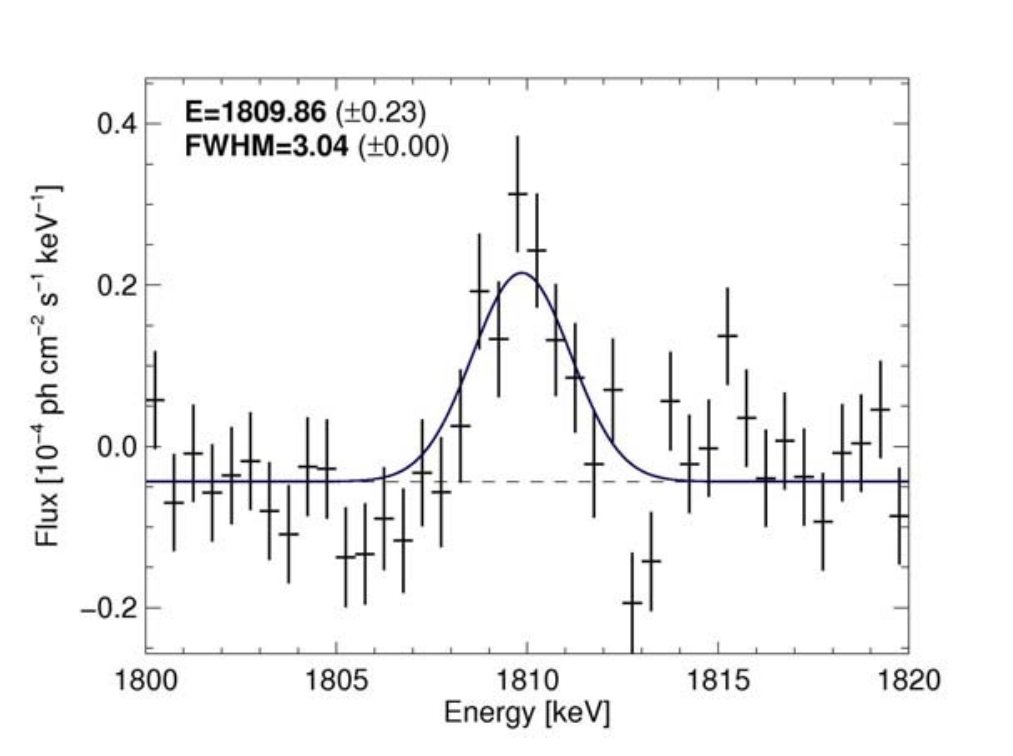}  
\caption{The \Al signal disentangled from the Scorpius-Centaurus region with INTEGRAL \cite{2010A&A...522A..51D}. }
\label{fig:26Al_sco-cen} 
\end{figure}   

Along the inner Galaxy, the nearby groups of massive stars associated with the Scorpius-Centaurus associations are at sufficiently-high galactic latitudes to be seen against the Galactic ridge $^{26}$Al emission (see Fig.~\ref{fig:26Al_sco-cen}) \cite{2010A&A...522A..51D}. 
The stellar association of Scorpius-Centaurus (Sco OB2) and its subgroups at a distance of about 100--150~pc \cite{1992A&A...262..258D, 1999AJ....117..354D, 1999AJ....117.2381P} apparently plays a major role, making up a significant part of the massive stars in the solar vicinity \cite{1999AJ....117..354D}. This association shows several subgroups of different ages (5, 16, and 17~My, with typical age uncertainties of 1--2 My; \cite{1989A&A...216...44D}, \cite{2008ApJ...688..377S}). The subgroup of Upper Sco has the right age for being dominant in terms of current massive-star mass ejection and supernova events (5~My, \cite{2009A&A...504..531V}). Signs of past supernova activity from this regions may, indirectly, be inferred from the morphology of the interstellar medium  \cite{1995SSRv...72..499F}, but also from estimates that the most massive star in Upper Sco presumably had $\sim 50\,M_\odot$ and thus may have exploded as a supernova about 1.5~Myr ago, and the pulsar PSR~J1932+1059 may be its compact remnant \cite{2000ApJ...544L.133H,2004ApJ...604..339C}. So, \Al with its characteristic My time scale can help to shed light on the nature of such suggested massive-star activity from this nearby group.
Stellar subgroups of different ages would result from a star forming region within a giant molecular cloud if the environmental effects of massive-star action of a first generation of stars (specifically shocks from winds and supernovae) would interact with nearby dense interstellar medium, in a scenario of  propagating or triggered star formation. Then later-generation ejecta would find the ISM pre-shaped by previous stellar generations. Such a scenario was proposed based on the different subgroups of the Scorpius-Centaurus Association \cite{1989A&A...216...44D, 2002AJ....124..404P} and the numerous stellar groups surrounding it \cite{2008A&A...480..735F}. 
Indications of recent star formation have been found in the L1688 cloud as part of the $\rho$~Oph molecular cloud, and may have been triggered by the winds and supernovae causing the \Al we observe. The young  $\rho$~Oph stars then could be interpreted as the latest signs of propagating star formation originally initiated from the oldest Sco-Cen subgroup in Upper Centaurus Lupus \cite{2008hsf2.book..351W}.  Many proposed scenarios of triggered star formation are only based on relatively weak evidence, such as the presence of Young Stellar Objects (YSOs) near shocks caused by massive stars.  
Positional evidence alone is not unequivocally considered to prove a triggered star formation scenario. Much more reliable conclusions can be drawn if the ages of the young stellar populations can be determined and compared to the moment in time at which an external shock from another star formation site arrived. Agreement of these timings would add convincing evidence for the triggered star formation scenario. 
These results are discussed in detail in \cite{2010A&A...522A..51D}.

\section{Summary and Prospects} 

The deepening of exposure resulting from the extended mission now gradually lifts details beyond the minimum thresholds of scientific significance, which could be important for improving insights into massive star evolution and their nucleosynthesis. SPI's capabilities to measure the gamma-ray lines associated with the decays of \Al, \Fe, and positron annihilations now combines with the modest spatial resolution to allow localized (i.e. region or star-group-specific) studies. These are more precise than global galactic averages, thus helping to constrain our uncertain models of massive star structure and evolution. Specifically, regions along the inner Galactic ridge, such as Scorpius Centaurus, Aquila, but also the Orion region, are promising laboratories for massive star studies with INTEGRAL in its later mission years. 

The European Science Foundation's ``EuroGenesis'' program (2010-2013) assembles several groups of scientists working on a varity of stellar-model, nuclear-reaction, and astronomical data collection aspects. This program also includes a special investigation of dust production from massive stars, adding a potential exploitation of infrared and meteoritic data for learning about massive star structure and evolution. It is one example which demonstrates how INTEGRAL measurements have now found their way into studies of general astrophysical interest and broad application.

{\bf Acknowledgements}
The INTEGRAL project of ESA and the SPI project has been completed under the responsibility and leadership of CNES/France.We are grateful to ASI, CEA, CNES, DLR, ESA, INTA, NASA and OSTC for support.
      The SPI anticoincidence system is supported by the German government through DLR grant 50.0G.9503.0.. The work for this paper was supported by the Munich Cluster of Excellence on ``Origins and Evolution of the Universe''.

%
\bibliographystyle{plain}
%


\begin{thebibliography}{10}

\bibitem{2004ApJ...604..339C}
S.~{Chatterjee}, J.~M. {Cordes}, W.~H.~T. {Vlemmings}, Z.~{Arzoumanian}, W.~M.
  {Goss}, and T.~J.~W. {Lazio}.
\newblock {Pulsar Parallaxes at 5 GHz with the Very Long Baseline Array}.
\newblock {\em \apj}, 604:339--345, March 2004.

\bibitem{1992A&A...262..258D}
E.~J. {de Geus}.
\newblock {Interactions of stars and interstellar matter in Scorpio Centaurus}.
\newblock {\em \aap}, 262:258--270, August 1992.

\bibitem{1989A&A...216...44D}
E.~J. {de Geus}, P.~T. {de Zeeuw}, and J.~{Lub}.
\newblock {Physical parameters of stars in the Scorpio-Centaurus OB
  association}.
\newblock {\em \aap}, 216:44--61, June 1989.

\bibitem{1999AJ....117..354D}
P.~T. {de Zeeuw}, R.~{Hoogerwerf}, J.~H.~J. {de Bruijne}, A.~G.~A. {Brown}, and
  A.~{Blaauw}.
\newblock {A HIPPARCOS Census of the Nearby OB Associations}.
\newblock {\em \aj}, 117:354--399, January 1999.

\bibitem{2006Natur.439...45D}
R.~{Diehl}, H.~{Halloin}, K.~{Kretschmer}, G.~G. {Lichti},
  V.~{Sch{\"o}nfelder}, A.~W. {Strong}, A.~{von Kienlin}, W.~{Wang}, P.~{Jean},
  J.~{Kn{\"o}dlseder}, J.-P. {Roques}, G.~{Weidenspointner}, S.~{Schanne},
  D.~H. {Hartmann}, C.~{Winkler}, and C.~{Wunderer}.
\newblock {Radioactive $^{26}$Al from massive stars in the Galaxy}.
\newblock {\em \nat}, 439:45--47, January 2006.

\bibitem{2010A&A...522A..51D}
R.~{Diehl}, M.~G. {Lang}, P.~{Martin}, H.~{Ohlendorf}, T.~{Preibisch},
  R.~{Voss}, P.~{Jean}, {J.-P.} {Roques}, P.~{von Ballmoos}, and W.~{Wang}.
\newblock {Radioactive $^{26}$Al from the Scorpius-Centaurus association}.
\newblock {\em \aap}, 522:A51+, November 2010.

\bibitem{2008A&A...480..735F}
D.~{Fern{\'a}ndez}, F.~{Figueras}, and J.~{Torra}.
\newblock {On the kinematic evolution of young local associations and the
  Scorpius-Centaurus complex}.
\newblock {\em \aap}, 480:735--751, March 2008.

\bibitem{1995SSRv...72..499F}
P.~C. {Frisch}.
\newblock {Characteristics of Nearby Interstellar Matter}.
\newblock {\em Space Science Reviews}, 72:499--592, May 1995.

\bibitem{2000ApJ...544L.133H}
R.~{Hoogerwerf}, J.~H.~J. {de Bruijne}, and P.~T. {de Zeeuw}.
\newblock {The Origin of Runaway Stars}.
\newblock {\em \apjl}, 544:L133--L136, December 2000.

\bibitem{2006ApJ...647..483L}
M.~{Limongi} and A.~{Chieffi}.
\newblock {The Nucleosynthesis of $^{26}$Al and $^{60}$Fe in Solar Metallicity
  Stars Extending in Mass from 11 to 120 M$_{\odot}$: The Hydrostatic and
  Explosive Contributions}.
\newblock {\em \apj}, 647:483--500, August 2006.

\bibitem{2009A&A...506..703M}
P.~{Martin}, J.~{Kn{\"o}dlseder}, R.~{Diehl}, and G.~{Meynet}.
\newblock {New estimates of the gamma-ray line emission of the Cygnus region
  from INTEGRAL/SPI observations}.
\newblock {\em \aap}, 506:703--710, November 2009.

\bibitem{2010A&A...511A..86M}
P.~{Martin}, J.~{Kn{\"o}dlseder}, G.~{Meynet}, and R.~{Diehl}.
\newblock {Predicted gamma-ray line emission from the Cygnus complex}.
\newblock {\em \aap}, 511:A86+, February 2010.

\bibitem{2002NewAR..46..535P}
S.~{Pl{\"u}schke}, M.~{Cervi{\~n}o}, R.~{Diehl}, K.~{Kretschmer}, D.~H.
  {Hartmann}, and J.~{Kn{\"o}dlseder}.
\newblock {Understanding $^{26}$Al Emission from Cygnus}.
\newblock {\em New Astronomy Review}, 46:535--539, July 2002.

\bibitem{2002AJ....124..404P}
T.~{Preibisch}, A.~G.~A. {Brown}, T.~{Bridges}, E.~{Guenther}, and
  H.~{Zinnecker}.
\newblock {Exploring the Full Stellar Population of the Upper Scorpius OB
  Association}.
\newblock {\em \aj}, 124:404--416, July 2002.

\bibitem{1999AJ....117.2381P}
T.~{Preibisch} and H.~{Zinnecker}.
\newblock {The History of Low-Mass Star Formation in the Upper Scorpius OB
  Association}.
\newblock {\em \aj}, 117:2381--2397, May 1999.

\bibitem{2008ApJ...688..377S}
C.~L. {Slesnick}, L.~A. {Hillenbrand}, and J.~M. {Carpenter}.
\newblock {A Large-Area Search for Low-Mass Objects in Upper Scorpius. II. Age
  and Mass Distributions}.
\newblock {\em \apj}, 688:377--397, November 2008.

\bibitem{2009A&A...504..531V}
R.~{Voss}, R.~{Diehl}, D.~H. {Hartmann}, M.~{Cervi{\~n}o}, J.~S. {Vink},
  G.~{Meynet}, M.~{Limongi}, and A.~{Chieffi}.
\newblock {Using population synthesis of massive stars to study the
  interstellar medium near OB associations}.
\newblock {\em \aap}, 504:531--542, September 2009.

\bibitem{2008hsf2.book..351W}
B.~A. {Wilking}, M.~{Gagn{\'e}}, and L.~E. {Allen}.
\newblock {\em {Star Formation in the {$\rho$} Ophiuchi Molecular Cloud}}, page
  351.
\newblock December 2008.

\end{thebibliography}

\end{document}